\documentclass[twocolumn]{aastex62}

\usepackage{amsmath,amssymb}
\usepackage{epstopdf}
\usepackage{ulem}
\usepackage{multirow}

\received{June 1, 2018}
\revised{June 7, 2018}
\accepted{\today}

\submitjournal{ApJ}

\shorttitle{Electron SSA at SNR shocks}
\shortauthors{Bohdan et al.}

\newcommand{\rev}{\textcolor{black}}

\newcommand{\mpon}{\textcolor{black}}
\newcommand{\mponn}{\textcolor{black}}
\newcommand{\mpo}{\textcolor{black}}
\newcommand{\ab}{\textcolor{black}}
\newcommand{\jn}{\textcolor{black}}
\newcommand{\art}{\textcolor{black}}
\newcommand{\ta}{\textcolor{black}}

\newcommand{\degree}{^{\rm o}}

\newcommand{\omci}{\Omega_\mathrm{i}}

\newcommand{\ms}{M_\mathrm{s}}
\newcommand{\ma}{M_\mathrm{A}}
\newcommand{\mi}{m_\mathrm{i}}
\newcommand{\me}{m_\mathrm{e}}
\newcommand{\lse}{\lambda_\mathrm{se}}
\newcommand{\lsi}{\lambda_\mathrm{si}}
\newcommand{\vsh}{v_\mathrm{sh}}

\begin{document}

\title{Kinetic \jn{simulations} of nonrelativistic perpendicular shocks of young supernova remnants. II. Influence of \art{shock-surfing acceleration} on downstream \jn{electron} spectra}

\correspondingauthor{Artem Bohdan}
\email{artem.bohdan@desy.de}

\author[0000-0002-5680-0766]{Artem Bohdan}
\affil{DESY, 15738 Zeuthen, Germany}

\author{Jacek Niemiec}
\affil{Institute of Nuclear Physics Polish Academy of Sciences, PL-31342 Krakow, Poland}

\author{Martin Pohl}
\affil{DESY, 15738 Zeuthen, Germany}
\affil{Institute of Physics and Astronomy, University of Potsdam, 14476 Potsdam, Germany}

\author{Yosuke Matsumoto}
\affil{Department of Physics, Chiba University, 1-33 Yayoi-cho, Inage-ku, Chiba 263-8522, Japan}

\author{Takanobu Amano}
\affil{Department of Earth and Planetary Science, the University of Tokyo, 7-3-1 Hongo, Bunkyo-ku, Tokyo 113-0033, Japan}

\author{Masahiro Hoshino}
\affil{Department of Earth and Planetary Science, the University of Tokyo, 7-3-1 Hongo, Bunkyo-ku, Tokyo 113-0033, Japan}

\begin{abstract}
\mpo{We explore electron pre-acceleration at high Mach-number nonrelativistic perpendicular shocks at, e.g., young supernova remnants, which are a prerequisite of further acceleration to very high energies via diffusive shock acceleration. Using fully kinetic particle-in-cell simulations of shocks and electron dynamics in them, we investigate the influence of shock-surfing acceleration at the shock foot on the nonthermal population of electrons downstream of the shock. The shock-surfing acceleration is followed by further energization at the shock ramp where the Weibel instability spawns a type of second-order Fermi acceleration.} 
The combination of these two processes leads to the formation of a nonthermal electron population, but the importance of shock-surfing acceleration becomes smaller for larger ion-to-electron mass ratio \mpo{in the simulation}. We discuss the resulting electron spectra and the relevance of our results to the physics of systems with real ion-to-electron mass ratio and fully three-dimensional \mpo{behavior}. 

\end{abstract}


\section{Introduction}\label{introduction}

Interaction of a supernova ejecta with an interstellar media results in high Mach number shocks which are often associated with a strong nonthermal emission. Electrons are significantly more radiative than ions and they produce nonthermal radiation in the radio band \citep{1953radi.book.....S}, in the X-ray band \citep{1995Natur.378..255K}, and in $\gamma$ rays \citep{1996A&A...307L..57P}. Is is commonly assumed that high energy electrons responsible for nonthermal emission are produced through diffusive shock acceleration \citep[DSA, e.g.,][]{1983RPPh...46..973D,1987PhR...154....1B}, also known as first-order Fermi acceleration process. During this process particles, which \jn{are} confined around the shock transition region by magnetic turbulence, experience repetitive head-on collisions with \jn{the} upstream and downstream plasma. However, DSA works only for particles \jn{whose} gyroradius is larger that the shock transition width, which in turn is defined by \mponn{the gyroradius} of the upstream ions.
Therefore some pre-acceleration or injection is needed to pick up particles from the thermal pool \jn{and} involve them in DSA. The electron injection process is notoriously more difficult than that for ions because of \mponn{the} large difference in mass \jn{of the two particle species}.

Here we study electron injection for conditions at young supernova remnant (SNR) shock waves, but the results of the current paper can be used for any high-Mach-number non-relativistic perpendicular shock, e.g. Saturn's bow shock~\citep{2016ApJ...826...48M}.
Present observational data do not to give clear constraints on the large-scale magnetic-field configuration in SNR shocks.
Different approaches at data modeling suggest either the presence of quasi-perpendicular magnetic field \citep{2009MNRAS.393.1034P,2010MNRAS.408..430S,2016A&A...587A.148W} or quasi-parallel configuration \citep{2004A&A...425..121R,2011A&A...531A.129B,2015MNRAS.449...88S} even for the same source. In this paper we continue our studies of perpendicular shocks ($\theta_{Bn}=90^o$) as the simplified form of the quasi-perpendicular case.

SNR shocks are characterized by high sonic, $\ms$, and Alfv\'enic, $\ma$, Mach numbers. In this regime the upstream ion kinetic energy can not be dissipated only via resistive (Joule) dissipation \citep{Marshall1955}, and the
simplest way of ion kinetic-energy dissipation is the reflection of a substantial part of the upcoming ions back upstream. Theoretical studies of perpendicular shocks \citep[e.g.,][]{1981GeoRL...8.1269L,1983PhFl...26.2742L}  demonstrate a great importance of the reflected ions \mponn{for} the structure of such shocks.
The  shock  transition  can  be  subdivided  into  a number of regions: an upstream with undisturbed plasma, a foot, a ramp, an overshoot, and the downstream region with shocked plasma.
The upstream ions are reflected by \mponn{the} shock potential at the ramp. 
Reflected ions interact with the incoming plasma and produce a number of instabilities in the shock foot. For our further discussion of electron pre-acceleration at high Mach number shocks two of these instabilities are of greatest importance –- the electrostatic Buneman instability \citep{1958PhRvL...1....8B} at the leading edge of the shock foot and the Weibel filamentation instability deeper in the shock foot \citep{1959PhFl....2..337F}, which further results in magnetic reconnection at the shock ramp \citep{2015Sci...347..974M}. 

A number of mechanisms are responsible for electron acceleration in the shock transition. Electrons can be accelerated during interaction with electrostatic waves resulting from \mpo{the} Buneman instability \citep{2000ApJ...543L..67S,2002ApJ...572..880H}. This process is also known as shock-surfing acceleration (SSA). Magnetic reconnection in the shock ramp results in acceleration of electrons via a number of channels discussed in \cite{2015Sci...347..974M}. In the turbulent ramp-overshoot region electrons undergo stochastic Fermi-like acceleration~\citep{2017ApJ...847...71B} or stochastic shock drift acceleration in the quasi-perpendicular case~\citep{Matsumoto2017}.

This paper is the second in \mponn{a} series of works that investigate different aspects of perpendicular shock physics by \jn{means} of fully kinetic simulations. High-resolution \jn{two dimensional (2D)} particle-in-cell (PIC) simulations are used to \jn{scrutinize} astrophysical shock physics for parameters that are close to those at young supernova remnants. Our numerical simulations represent a large enough portion of the shock surface to demonstrate in detail shock physics and variety of electron acceleration processes.

In the previous paper (\cite{2019ApJ...878....5B}, hereafter Paper I) we discussed electron pre-acceleration at the leading edge of the shock foot via SSA process. We showed that in high-$\ma$ shock simulations the strength of the electrostatic wave modes in the shock foot is determined by the Alfv\'enic Mach number in relation to the trapping condition~\citep{2012ApJ...755..109M}:
\begin{equation}
  \ma \geq (1+\alpha)  \left(\frac{m_\mathrm{i}}{m_\mathrm{e}}\right)^{\frac{2}{3}},
  \label{trapping}
\end{equation}
\jn{where $\alpha$ is the fraction of shock-reflected ions, and $m_\mathrm{i}$ and $m_\mathrm{e}$ denote the ion and electron mass, respectively.}
The more $\ma$ exceeds \mponn{this} trapping \jn{limit}, the stronger \jn{is} the intensity of the Buneman waves. Shocks with Alfv\'enic Mach numbers satisfying the trapping condition demonstrate similar wave strengths in simulations \mponn{with} different ion-to-electron mass ratios, \rev{$m_i/m_e$, hereafter referred to as the mass ratio}. 

\jn{Two-dimensional simulations of perpendicular shocks need to specify the magnetic field orientation with respect to the simulation plane. This choice has influence on the shock physics observed in the simulations. We have shown that shocks}
in simulations with in-plane magnetic field demonstrate electrostatic wave intensities lower than those observed in the out-of-plane case, even if the modified trapping condition~\citep{2017ApJ...847...71B} is satisfied:
\begin{equation}
 \ma \geq \sqrt\frac{2}{1+\sin^2\varphi}
 (1+\alpha)  \left(\frac{m_\mathrm{i}}{m_\mathrm{e}}\right)^{\frac{2}{3}},
\label{trappingnew}
\end{equation}
which was introduced to compensate \mponn{the} out-of-plane motion of reflected ions in simulations with \jn{an} in-plane magnetic-field \jn{component, defined through the angle $\varphi$ between the simulation plane and the perpendicular magnetic field}.

The SSA mechanism always produces larger fractions of pre-accelerated electrons in simulations with out-of-plane configuration, even if the intensities of the Buneman waves are similar as in the in-plane case because of almost twice larger velocity difference between reflected ions and upstream electrons in \jn{the} out-of-plane case. In Paper I we concluded that the same SSA efficiency as in the out-of-plane or 3D case can not be achieved \mponn{with} in-plane simulations. Thus to reproduce the realistic 3D shock physics \mponn{a} combination of in-plane and out-of-plane simulations are needed.

The main goal of this work is to \jn{scrutinize the} influence of non-adiabatic acceleration processes in the shock transition on \mponn{the population of electrons} \jn{that have been already} pre-accelerated via SSA. Previously we demonstrated \citep{2017ApJ...847...71B,Matsumoto2017} that simulations with in-plane magnetic-field configuration can be used as a good approximation of 3D perpendicular shock physics in the shock foot (beyond the Buneman wave zone with SSA), \jn{the} ramp and \jn{the} overshoot regions. Processes \jn{that} take place in the shock transition might change the ratio between thermal and nonthermal \jn{electrons} pre-accelerated by SSA. This study is based on the same simulations as in Paper I and complement our previous investigations of nonrelativistic perpendicular shocks \citep[e.g.,][]{2012ApJ...755..109M,2013PhRvL.111u5003M,2015Sci...347..974M,2016ApJ...820...62W,2017ApJ...847...71B}.

The present paper is structured as follows. We present a short description of the simulation setup in Section~\ref{sec:setup}. The results are presented in Section~\ref{results}. Section~\ref{summary} contains the summary and discussion.

\section{Simulation Setup} \label{sec:setup}

The relativistic electromagnetic two-dimensional PIC code is used to examine the SNR shock physics. This is a 2D3V-adapted and modified version of the TRISTAN \citep{Buneman1993} code with MPI-based parallelization \citep{2008ApJ...684.1174N,2016ApJ...820...62W} and the option to trace individual particles. 

\begin{figure}[htb]
\centering
\includegraphics[width=\linewidth]{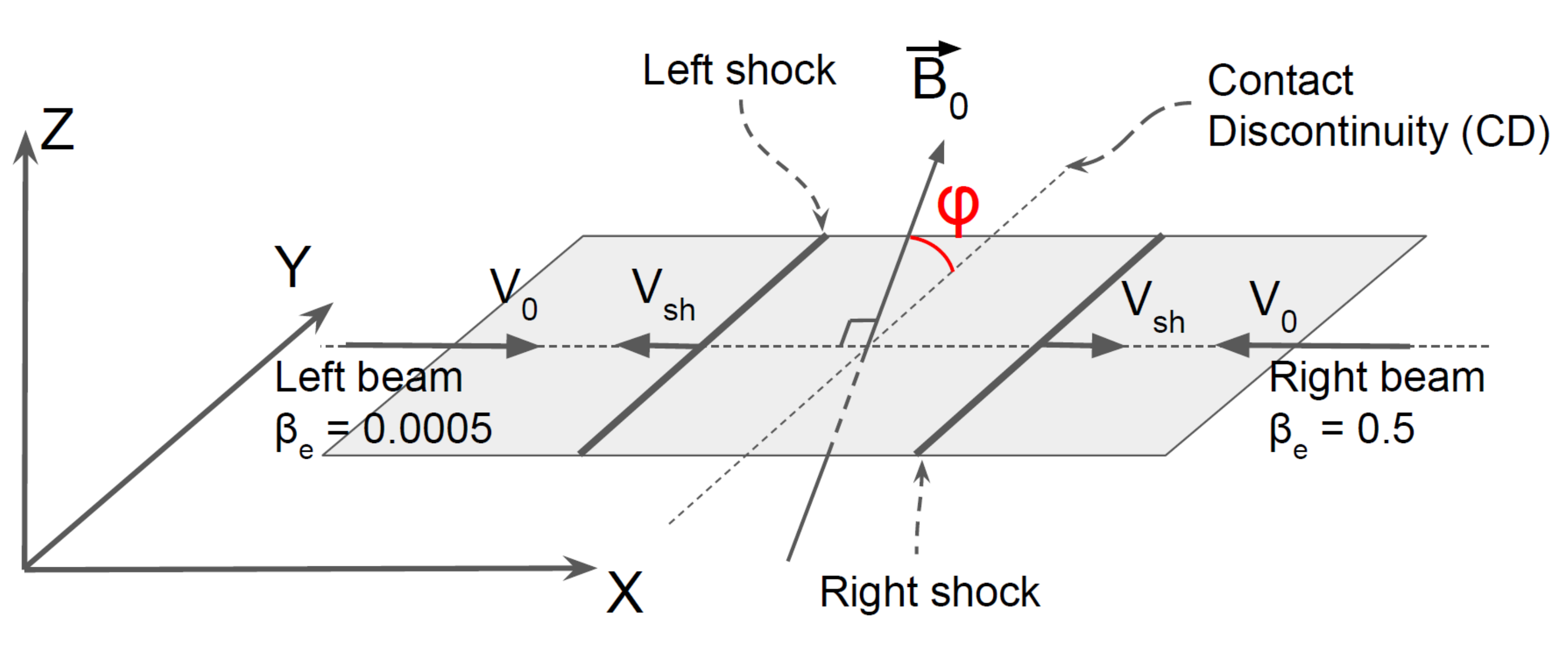}
\caption{Illustration of the simulation setup.} 
\label{setup}
\end{figure}

The simulation setup used in this work is the flow-flow method used in Paper I and in \cite{2017ApJ...847...71B}. An illustration of the simulation setup is presented in Figure~\ref{setup}. It considers an interaction of two counterstreaming electron-ion plasma flows. As a result of the two plasma slabs collision two shocks are formed propagating in opposite directions and separated by a contact discontinuity (CD). This setup offers more freedom in the choice of physical parameters because two shocks in plasma environments with different parameters can be investigated at the same time. Here we refer to shocks as the \emph{left} \jn{(L)} and the \emph{right} \jn{(R)} shocks. The absolute values of the beam velocities are equal, $v_{\rm L}=v_{\rm R}=0.2c$.

   \begin{table*}[!t]
      \caption{Simulation Parameters}
         \label{table-param}
     $$ 
\begin{array}{p{0.07\linewidth}ccrcccccr}
\hline
\hline
\noalign{\smallskip}
Runs  & \mi/\me &   \ma & \multicolumn{2}{c}{\ms} & \multicolumn{2}{c}{\beta_{\rm e}} & Eq.~\ref{trapping} & \multicolumn{2}{c}{Eq.~\ref{trappingnew}}\\
 & & & ^*1 & ^*2 & ^*1 & ^*2 & \alpha=0.2  & \alpha=0.2 & (0.5) \\
\noalign{\smallskip}
\hline
\noalign{\smallskip}
A1, A2   & 50  &  22.6  & 1104 & 35 & 5 \cdot 10^{-4} & 0.5 & 16  & 22.4 & (28) \\
B1, B2    & 100 &   31.8  & 1550 & 49 & 5 \cdot 10^{-4} & 0.5 & 26  &  36 & (46)\\
C1, C2   & 100 &   46  & 2242 & 71 & 5 \cdot 10^{-4} & 0.5 & 26  &  36 & (46)\\
D1, D2   & 200 &   32  & 1550 & 49 & 5 \cdot 10^{-4} & 0.5 & 41  &  58 & (72)\\
E1, E2   & 200 &   44.9  & 2191 & 69 & 5 \cdot 10^{-4} & 0.5 & 41  &  58 & (72) \\
F1, F2   & 400 &  68.7  & 3353 & 106 & 5 \cdot 10^{-4} & 0.5 & 65  &  92 & (115)\\
\noalign{\smallskip}
\hline
\end{array}
     $$ 
\tablecomments{Parameters of simulation runs described in this paper. Listed are: the ion-to-electron mass ratio $\mi/\me$, and Alfv\'enic  and sonic Mach numbers, $\ma$ and $\ms$, the latter separately for the \emph{left} (runs *1) and the \emph{right} (runs *2) shock. We also list the electron plasma beta, $\beta_{\rm e}$, for each simulated shock and the critical Alfv\'enic Mach number (Eq.~\ref{trapping}) for $\alpha=0.2$, as well as the modified trapping condition (Eq.~\ref{trappingnew}) calculated for $\alpha=0.2$ and  $\alpha=0.5$ (in brackets).
All runs use the in-plane magnetic field configuration, $\varphi=0^o$.} 
   \end{table*}

Plasma beams have equal density but different temperatures. The temperature between the two beams differs by factor of $1000$. Thus the \emph{electron} plasma beta (the ratio of the electron plasma pressure to the magnetic pressure) for the left beam is $\beta_{\rm e,L}=5 \cdot 10^{-4}$ and $\beta_{\rm e,R}=0.5$ for the right beam. The \emph{sonic} Mach number, $M_{\rm s}$, of the two shocks differ by a factor of $\sqrt{1000} \simeq 30$.

The large scale magnetic field makes \jn{the} angle, $\varphi$, with the simulation plane. \jn{Here} all runs assume the in-plane magnetic field configuration, $\varphi=0^o$. Taking into account the adiabatic index $\Gamma=5/3$ in \jn{such} field configuration, the resulting expected shock speeds take values $\vsh=0.263c$ in the \emph{upstream} reference frame. The Alfv\'en velocity is defined as $v_{\rm A}=B_{\rm 0}/\sqrt{\mu_{\rm 0}(N_e\me+N_i\mi)}$, where $\mu_{\rm 0}$ is the vacuum permeability, $N_i$ and $N_e$ are the ion and the electron number densities, and $B_0$ is the far-upstream magnetic-field strength. The sound speed reads $c_{\rm s}=(\Gamma k_BT_{\rm i}/\mi)^{1/2}$, where $k_B$ is the Boltzmann constant and $T_{\rm i}$ is the ion temperature defined as $k_BT_i=m_i v_{\rm th,i}^2/2$ ($v_{\rm th,i}$ is defined as the most probable speed of the upstream plasma \jn{ions} in the upstream reference frame).
The Alfv\'enic, $\ma=\vsh/v_{\rm A}$, and sonic, $\ms=\vsh/c_{\rm s}$, Mach numbers of the shocks are defined in the conventional \emph{upstream} reference frame.

The weakly magnetized plasmas are considered here. The ratio of the electron plasma frequency, $\omega_{\rm pe}=\sqrt{e^2N_e/\epsilon_0\me}$, to the electron gyrofrequency, $\Omega_{\rm e}=eB_0/\me$, \jn{is} in the range $\omega_{\rm pe}/\Omega_{\rm e}=8.5-17.3$. Here, $e$ is the electron charge, and $\epsilon_0$ is the vacuum permittivity.

The electron skin depth in the upstream plasma is common for all runs and equals $\lse=20\Delta$, where $\Delta$ is the size of grid cells. As the unit of length the ion skin depth, $\lsi=\sqrt{\mi/\me}\lse$, is used.
The time-step we use is $\delta t=1/40\,\omega_{\rm pe}^{-1}$. 
The time-scales are preferably \jn{given} in terms of the upstream ion Larmor frequency, $\Omega_{\rm i}$, where $\Omega_{\rm i}=eB_0/\mi$. \rev{The simulation time is typically $t=(6-8)\Omega_{\rm i}^{-1}$.} Number density in the far-upstream is 20 particle per cell for each species.
For more detailed description of the simulation setup and definition of the shock parameters see Paper~I.

Here we discuss results of six  large-scale numerical experiments (runs A--F), that feature in total twelve simulated shocks. Here we refer to each of these shock cases as to a separate simulation run, and label the shocks in the left plasma ($\beta_{\rm e,L}=5 \cdot 10^{-4}$) with *1, and the right shocks with *2 ($\beta_{\rm e,R}=0.5$).
The derived parameters of the simulation runs described in this paper are listed in Table~\ref{table-param}.

The simulations runs cover a wide range of mass ratios and Alfv\'enic Mach numbers, which permits an investigation of the influence of these parameters on the electron acceleration efficiency and to scale our results to the realistic mass ratio. Note, that some aspects of the shock physics, namely \jn{the} SSA efficiency, in all runs have been \jn{already} discussed in Paper~I. 

The shock parameters are chosen to investigate the influence of SSA process on \jn{the} formation of nonthermal electron populations in the shock downstream. Runs A, B, E, and F fulfill the trapping condition of Equation~\ref{trapping}, run C fulfills the modified trapping condition of Equation~\ref{trappingnew} \art{with $\alpha=0.5$} and the Alfv\'enic Mach number in case of run D is below the trapping limit (Eq.~\ref{trapping}). \mponn{Run A marginally satisfies the modified trapping condition for $\alpha=0.2$. However, as discussed in \cite{2017ApJ...847...71B}, the fraction of reflected ions is larger then 0.2 in the in-plane case, and the Alfv\'enic Mach number of run A does not satisfy the modified trapping condition of Equation~\ref{trappingnew}, if $\alpha$ is calculated directly from the simulation. Thus we keep denoting run A as satisfying the trapping condition of Equation~\ref{trapping}.}

These different cases are characterized by different SSA efficiency and \jn{different} number of pre-accelerated electrons at the leading edge of the shock foot. 
\art{The \jn{contribution of the} latter to \mponn{the} nonthermal electron population \jn{in} the downstream region might depend on the Alfv\'enic Mach number \mponn{and the} mass ratio.} The goal of this paper is to clarify these dependencies.

\section{Results} \label{results}


We begin the presentation of \jn{our} results with a description of the downstream spectra for all runs in Section~\ref{downstream}. We discuss trends observed \jn{for parameters that characterize the} downstream spectra.
In Section~\ref{without-BI} we describe the formation of nonthermal electron populations during interaction of electrons with Buneman and Weibel instabilities.  
Then in Section~\ref{trans-coef} we discuss which portion \jn{of electrons} pre-accelerated through SSA can reach the nonthermal tail \jn{in the downstream spectra}. 

\subsection{Downstream Electron Spectra} \label{downstream}

\begin{figure}[!t]
\centering
\includegraphics[width=0.95\linewidth]{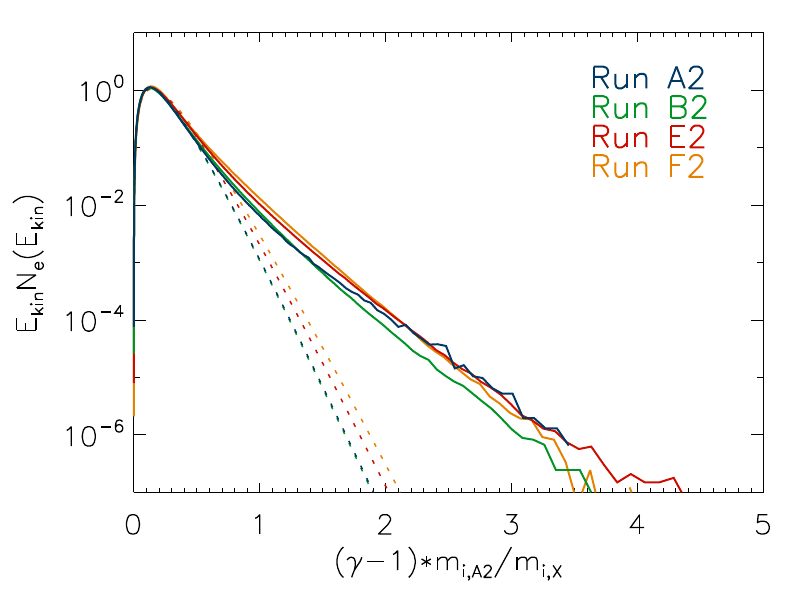}
\caption{\mpo{Rescaled electron spectra in the downstream region of shocks with $\beta_{e}=0.5$. \rev{The $x$-axis is corrected by factor $m_{\rm i,A2}/m_{\rm i,X}$ for each spectra, where $m_{\rm i,A2}=50$ is the ion mass for run A2, and $m_{\rm i,X}=50,100,200,400$ the ion mass for runs A2-F2 shown in the figure.} Dashed lines represent fits of a relativistic Maxwellian to the low-energy part of the spectra. The line color identifies the run.}}
\label{spectra4}
\end{figure}

\begin{figure}[!t]
\centering
\includegraphics[width=0.95\linewidth]{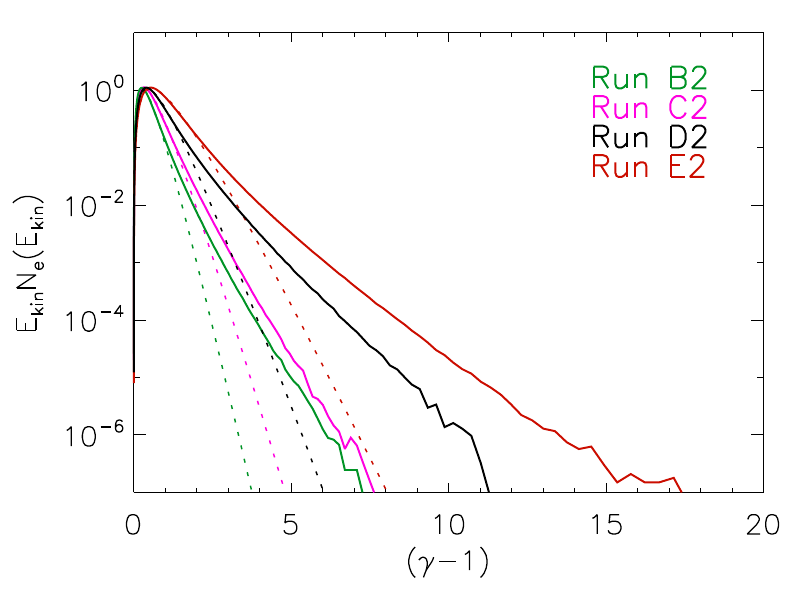}
\caption{Electron spectra as in Figure~\ref{spectra4}, but without rescaling in energy and for runs B2, C2, E2 and D2. \rev{Dashed lines represent fits of a relativistic Maxwellian to the low-energy part of the spectra.}}
\label{spectra}
\end{figure}

   \begin{table*}
      \caption{\mpo{Spectral Parameters for the Downstream Region}}
         \label{table-spectra2}
\centering
\begin{tabular}{lcccccc}
\hline
\hline
\noalign{\smallskip}
Run   & $\mi/\me$ & $N_{e,BI}/N_{e,tot}$ (\%)  & NTEF (\%) &  $k_B T/m_ec^2$ &  $k_B T_\mathrm{RH}/m_ec^2$ & max($E_{\rm kin}/m_ec^2$) \\
\noalign{\smallskip}
\hline
\noalign{\smallskip}
A1  &  50  & 0.43   &  $0.1\pm0.03$  & $0.107\pm0.004$  & 0.325    & $2.04\pm0.27$ \\
A2  &  50  & 0.43   &  $0.28\pm0.08$  & $0.091\pm0.004$  & 0.325   & $1.9\pm0.21$  \\
\noalign{\smallskip}
\hline
\noalign{\smallskip}
B1  &  100  & 0.46   &  $0.17\pm0.03$  &  $0.216\pm0.004$  & 0.65   &  $5.67\pm0.54$ \\
B2  &  100  & 0.46   &  $0.55\pm0.08$  &  $0.183\pm0.004$  & 0.65  &  $5.5\pm0.5$   \\
C1  &  100  & 0.6   &  $0.23\pm0.03$  &  $0.253\pm0.002$  & 0.65  &  $6.05\pm0.58$ \\
C2  &  100  & 0.6   &  $0.36\pm0.05$  &  $0.217\pm0.003$  & 0.65  &  $5.25\pm0.47$  \\
\noalign{\smallskip}
\hline
\noalign{\smallskip}
D1  &  200 & 0.34  &  $0.4\pm0.2$  &   $0.332\pm0.024$  & 1.3   & $8.6\pm0.8$  \\
D2  &  200 & 0.34  &  $0.7\pm0.07$ &   $0.28\pm0.003$  & 1.3   & $8.34\pm0.73$  \\
E1  &  200 & 0.49   &  $0.17\pm0.03$  &   $0.394\pm0.005$  & 1.3  &  $10.3\pm0.97$   \\
E2  &  200 & 0.49  &  $0.56\pm0.05$  &   $0.368\pm0.009$  & 1.3  &  $11.6\pm1.0$  \\
\noalign{\smallskip}
\hline
\noalign{\smallskip}
F1  &  400 & 0.44   &  $0.46\pm0.05$  &   $0.765\pm0.035$  & 2.6   & $22.7\pm1.8$ \\
F2  &  400 & 0.44   &  $0.57\pm0.07$  &   $0.732\pm0.02$  & 2.6  & $22.8\pm1.8$   \\
\noalign{\smallskip}
\hline
\end{tabular}
\smallskip
\tablecomments{\rev{$N_{e,BI}/N_{e,tot}$ is the fraction of electrons preacclerated via SSA in the Buneman instability region (see Paper I).} NTEF is nonthermal electron fraction in the downstream spectra. \rev{ $k_B T/m_ec^2$ is the downstream electron temperature, $k_B T_\mathrm{RH}/m_ec^2$ is the downstream electron temperature expected from the Rankine-Hugoniot jump conditions (Eq.~\ref{eq:enerT}), and max($E_{\rm kin}/m_ec^2$) is the kinetic energy of 1000 most energetic electrons residing downstream.} } 
   \end{table*}

In this section we discuss the processes that \mpo{provide} nonthermal electrons in the final electron spectra observed downstream of high-Mach-number shocks. As discussed in \cite{2017ApJ...847...71B}, the downstream spectrum in simulations with out-of-plane magnetic field is just the adiabatically compressed electron distribution generated in the Buneman instability region. Hence, we focus here on the acceleration processes occurring in simulations with in-plane field configurations. Figures~\ref{spectra4} and~\ref{spectra} show electron spectra downstream of \emph{right} shocks (runs A2-F2), propagating in moderate-temperature plasmas with $\beta_{\rm e}=0.5$, and Table~\ref{table-spectra2} lists parameters of downstream spectra for all simulated shocks with $\varphi=0\degree$. 
Since the shock self-reformation causes quasi-periodic variations in the electron acceleration efficiency, the resulting nonthermal electron fractions have a nonuniform spatial distribution behind the shock \citep{2017ApJ...847...71B}. We account for these spectral distortions by averaging over a region downstream of the overshoot that contains particles produced over at least two cycles of shock reformation, i.e., of length $2\times 1.55\,\omci^{-1}\times \vsh\simeq 25\lsi$. Here the factor $1.55\omci^{-1}$ is the average time-scale of shock reformation. \rev{Electrons for the spectra are selected at times close to the end of simulations.}

The nonthermal electron fraction \art{(NTEF)} listed in Table~\ref{table-spectra2} is calculated \mpo{as excess over} the Maxwellian fits to the low-energy part of the \art{downstream} spectra. \art{The temperature, $k_B T/m_ec^2$, is \mponn{derived by fitting a} relativistic Maxwellian.} The maximum kinetic energy of downstream electrons is estimated as the average energy of the 1000 most energetic electrons residing downstream.   

Figure~\ref{spectra4} shows electron spectra for the simulation runs A2, B2, E2 and F2, for which the Alfv\'enic Mach number satisfies the trapping condition (Eq.~\ref{trapping}). The spectra are scaled \jn{with} the \rev{mass ratio}. \mpo{We \ab{have} demonstrated in \ab{Paper I} that in all these cases the same fraction, $\sim 0.45\%$ \rev{(see Table~\ref{table-spectra2})}, of \art{SSA-pre-accelerated electrons} was observed in the Buneman wave region. To be noted from Figure ~\ref{spectra4} is that the same is true for the downstream spectra. One might \jn{thus} be inclined to conclude} 
that the NTEF is determined by the efficiency of pre-acceleration through the SSA process. However, as we demonstrate below, such a conjecture 
cannot be maintained.

Runs B and C probe different Alfv\'enic Mach numbers for the same mass ratio $\mi/\me=100$. In particular, $\ma$ \mpo{in run C exceeds the modified trapping limit} (Eq.~\ref{trappingnew}), and so in this case stronger electrostatic waves are observed in the shock foot, producing  larger numbers of \art{SSA-pre-accelerated electrons} \ab{($\sim 0.6\%$)}, as we demonstrated for right shocks in \ab{Paper I}. 
However, \mpo{this trend does not persist to the downstream region where the higher-$\ma$ shocks show similar (cold plasmas, runs B1 and C1) or smaller (warm plasmas, runs B2 and C2) fractions of nonthermal electrons than shocks with} lower Mach number (see Table~\ref{table-spectra2}).  
\mpo{The same counter-intuitive behavior is observed in simulations with $\mi/\me=200$ (runs D and E). The Alfv\'enic Mach number in run D is significantly below the trapping limit, and SSA is \art{relatively} inefficient \ab{with fraction of \art{SSA-pre-accelerated electrons} about $\sim 0.35\%$}. Nevertheless, the observed NTEFs are \ab{even} larger than those for shocks with higher $\ma$. Clearly, electron injection at high-Mach-number shocks is not entirely determined by SSA \ab{at the leading edge of} the shock foot. Our results suggest that other processes are at play that pre-accelerate and shape the electron spectra in the downstream region.}   

Comparing shocks in plasmas with different plasma beta, but otherwise similar conditions, one finds that shocks in cold plasmas ($\beta_{\rm e}=5 \cdot 10^{-4}$) have a slightly higher downstream temperature and a somewhat smaller NTEF than the $\beta_\mathrm{e}=0.5$ shocks, whereas the maximum kinetic energy is comparable. 
\mpo{The latter suggests that the processes of energy transfer work with comparable efficiency at shocks in cold and warm plasmas, and that a stronger randomization of energy at low-$\beta_{\rm e}$ shocks increases the temperature and hence lowers the excess of nonthermal particles. Certainly} the mechanisms of heating and pre-acceleration cannot be simply separated. A detailed scrutiny of the heating processes is outside the scope of this work and will be addressed in the future. Here we only note the trends observed that the electron temperature grows with the Mach number of the shock (compare runs B, C and D, E) and the mass ratio (see runs B, D and C, E).    

\mpo{If the Rankine-Hugoniot jump conditions \jn{are} applied \art{and ion-electron energy equipartition is assumed}, the normalized downstream temperature would be
\begin{equation}
\frac{k_B T_\mathrm{RH}}{\me c^2}=\frac{1}{2}\, \frac{3}{16}\,\frac{\mi}{\me}\,\frac{\vsh^2}{c^2}\simeq  6.5\cdot 10^{-3}\, \frac{\mi}{\me}\ .
\label{eq:enerT}
\end{equation}
\rev{Values of the theoretically expected downstream temperatures are listed in Table~\ref{table-spectra2}.}
Inspection of \rev{this table} indicates that the actual electron temperatures in our simulations \rev{are a factor of $3.5\pm1$}
smaller than the hydrodynamical expectation. This is not surprising, as we model a collisionless shock, and is actually in line with the observed temperature equilibration at fast SNR shocks \citep{2001ApJ...547..995G,2005AdSpR..35.1017R}. } 

\subsection{Where Are High-Energy Electrons Produced?} \label{without-BI}

\begin{figure*}[!t]
\centering
\includegraphics[width=0.95\linewidth]{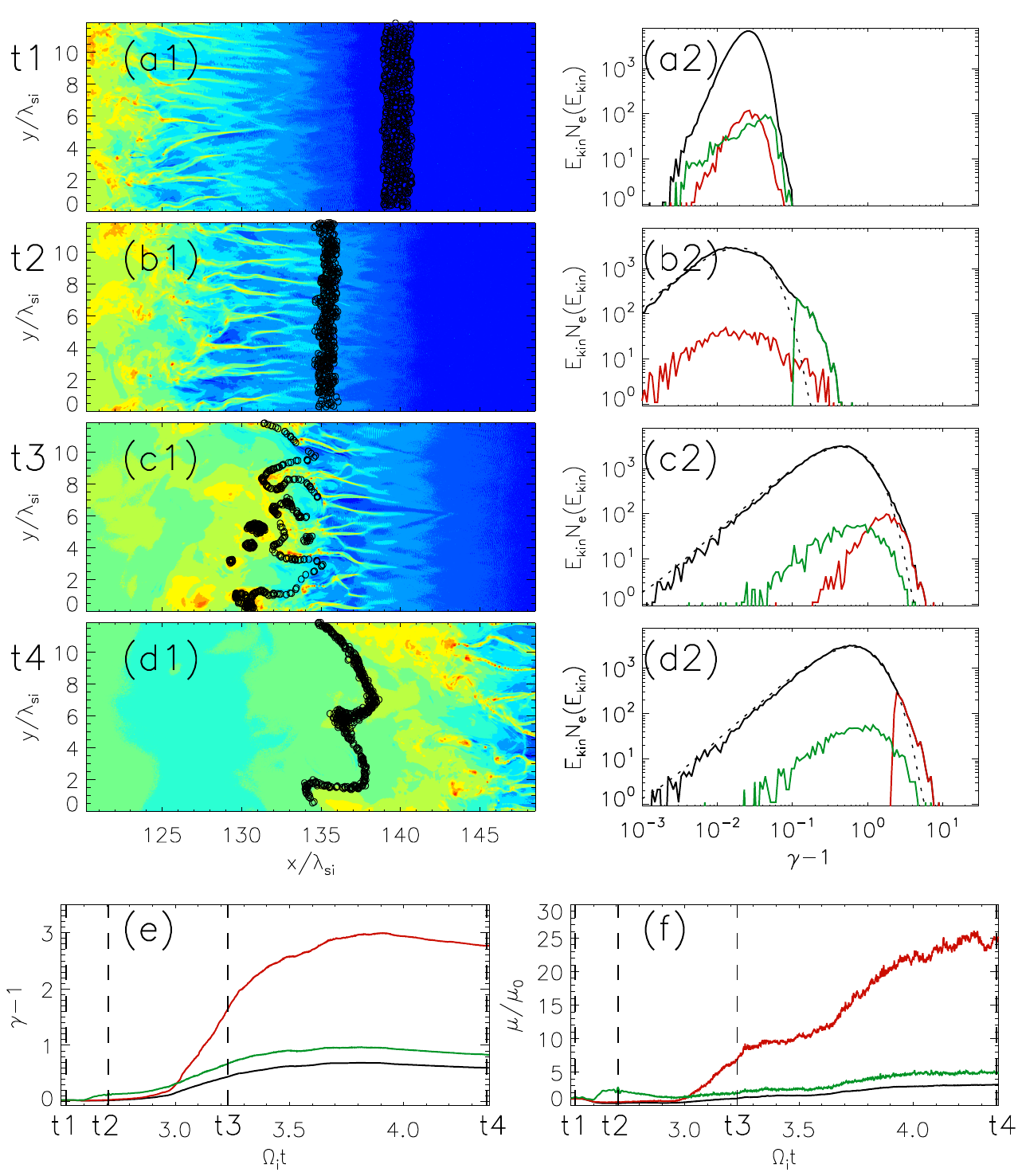}
\caption{\mpo{Evolution of the energy and the magnetic moment of traced electrons \rev{for run E2}. Panels (a1), (b1), (c1) and (d1) present the positions of traced electrons (black dots) at the time t1, t2, t3 and t4 on density maps of the shock region. Panels (e) and (f) display the evolution of the average electron energy and the normalized average magnetic \jn{moment}, correspondingly. The dashed lines in panels (e) and (f) are time markers for t1-t4. Panels (b2), (c2) and (d2) show electron spectra compared to fits of a relativistic Maxwellian (dashed line), \rev{where the bulk energy is negligible compared to the thermal energy of electrons}. 
Black lines in panels (a2)-(d2), (e) and (f) refer to all traced electrons, green lines correspond to electrons energized in the Buneman zone, and red lines are for high-energy electrons downstream of the shock.
}
}
\label{ele_acc_HL1_fig}
\end{figure*}

In \citet{2017ApJ...847...71B} we \mpo{demonstrated} that for configurations with in-plane magnetic field electrons accelerated through SSA are further energized by non-adiabatic processes in the shock ramp and at the overshoot, where they undergo stochastic second-order Fermi acceleration in interactions with magnetic turbulence. However, the contribution of SSA to the final downstream spectra has not been discussed, and \mpo{we have not quantified} the number of highly energetic electrons that would experience only scattering in the shock without interactions with the Buneman waves. In this section we address these issues to \mpo{establish the impact of SSA on the generation of nonthermal electrons and where in the shock nonthermal electrons are produced.}

Figure~\ref{ele_acc_HL1_fig} illustrates the \mpo{evolution of the spectra of a selection of electrons crossing} the shock. Our analysis is based on tracing data of about $5\cdot 10^5$ individual particles collected for run E2. The spectra are presented at four \mpo{points in time in panels (a2)--(d2), for which the locations of the particles are marked in} density maps in panels (a1)--(d1). The spectra for all traced particles are plotted with black solid lines. \mpo{Beginning at time $t2$ the bulk motion of electrons is negligible, and the spectra in panels (b2)--(d2) include a best-fit} relativistic Maxwellian indicated with dashed lines. Two subsets of electrons are chosen for the analysis. The first subset is defined at time $t_2$ and consists of electrons that have been pre-accelerated \mpo{by SSA} in the Buneman-instability region and populate the nonthermal tail in the spectrum. For this subset we chose particles with kinetic energies higher than $(\gamma-1)>0.1$. Their distributions in Figure~\ref{ele_acc_HL1_fig}(a2)--(d2) are shown with green solid lines, and we refer to these particles as to the Buneman instability (BI) electrons. The second subset \mpo{comprises} electrons that at time $t_4$ have energies $(\gamma-1) \geq 5k_BT/\me c^2$, where $T$ is the downstream temperature derived for the distribution of traced electrons. 
The limiting energy $5\cdot k_BT/\me c^2$ is chosen here \emph{ad-hoc}, as an approximate energy above which an 
\ab{excess of electron over the Maxwellian fit is observed.}
\art{Note that the change of the limiting energy to a lower or a higher value does not change the main results of this paper.}
In the following we refer to electrons in this subset as to the high-energy (HE) electrons. Their distributions are marked with red solid lines in Figure~\ref{ele_acc_HL1_fig}(a2)--(d2). Note, that the two sets of BI and HE electrons \mpo{have overlap}. In particular, if SSA were the first-stage acceleration for all electrons, the intersection of the two sets would be 100\%. Any smaller intersection suggests that the nonthermal population of electrons is determined not only by SSA. Panels (e) and (f) in Figure~\ref{ele_acc_HL1_fig} show the time evolution of the average energy and magnetic moment, respectively, for all selected electron sets.       

As it is impossible to distinguish \mpo{thermal and nonthermal particles on the basis of the spectrum alone, we need to look for a deviation of the observed number of high-energy electrons from that expected for a thermal distribution.} For a Maxwellian energy distribution with temperature $k_BT$, \mpo{about 1.85\% of the particles have energies exceeding} $5k_BT$, and we
define \ab{the NTEF as a difference between the percentage of high-energy electrons ($N_{\rm HE}(E>5k_BT)$)} and the share of thermal particles in the tail, \ab{${\rm NTEF}=N_{\rm HE}-0.0185$}.  

At time $t_1$ the particles are upstream of the Buneman wave region and their distribution is a drifting Maxwellian with a drift velocity $v_0$ (Fig.~\ref{ele_acc_HL1_fig}a1--a2). The energy distribution of electrons that would later become HE electrons is consistent with thermal. \mpo{In contrast, the BI electrons (green line) appear to have a large random velocity, appearing here as excess at energies lower and higher than the \mpo{bulk-flow energy}, $\gamma_0-1$.} Our tracing analysis of BI electrons shows that particles with \mpo{fast random motion} are indeed preferably accelerated via SSA.

After their interaction with the Buneman waves, BI electrons \mpo{by definition of the sample populate the high-energy tail of the spectrum} (Fig.~\ref{ele_acc_HL1_fig}b1--b2, see above). Their average energy, $\langle\gamma-1\rangle\simeq 0.15$ (Fig.~\ref{ele_acc_HL1_fig}e at $t_2$), is approximately seven times higher \mpo{than that of all electrons, and their average magnetic moment has more than doubled} (Fig.~\ref{ele_acc_HL1_fig}f).  
The distribution of HE electrons is still approximately thermal, though a slight increase in the average energy of $\sim 30\%$ \mpo{can be observed. This marginal energization does not seem to be non-adiabatic, as the average} magnetic moment remains constant. 

At time $t_3$ (Fig.~\ref{ele_acc_HL1_fig}c1--c2) the electrons are deep in the Weibel instability region and at the overshot. \mpo{The bulk of electrons is still thermal, but the temperature has increased by about a factor $25$. BI electrons received little energy boost, and their average energy is only $\sim 55\%$ larger than that of all traced electrons. In fact, after time $t_2$ the energy increment of BI electrons 
is virtually the same as that of all electrons, indicating a bulk heating and no acceleration, which is in line with the approximate} constancy of the magnetic moment. \mpo{In contrast, around} $t\sim 3/\omci$ the energy of HE electrons grows considerably faster, and they \mpo{dominate the high-energy tail of the spectrum.} The HE electrons \mpo{have an average energy about $ 3.5$ times higher than that of all traced electrons, on account of non-adiabatic second-order Fermi-like acceleration processes \citep{2017ApJ...847...71B} that are also documented by a large increase in the average} magnetic moment  (Fig.~\ref{ele_acc_HL1_fig}f).

At time $t_4$ the traced electrons reside in the downstream region (Fig.~\ref{ele_acc_HL1_fig}d1--d2).
BI electrons remain thermal with a temperature only $40\%$ higher than \mpo{that of the bulk. By definition, the HE electrons form the high-energy tail of the spectrum. We deduce from the evolution of the energy and magnetic moment (Fig.~\ref{ele_acc_HL1_fig}e--f), that the energy gain of HE electrons arises from both compression and non-adiabatic acceleration processes. BI electrons in general contribute little to the high-energy spectral tail (see below)}.

\mpo{Note, that the downstream nonthermal fraction cited for run E2 in Table~\ref{table-spectra2} is 0.56\% higher than that of the HE electrons (0.35\%), which reflects the temporal variation in electron energization imposed by shock reformation. The number in the table is calculated for a large part of the downstream region and hence represents a proper average, as do the values quoted in \citet{2017ApJ...847...71B}.}


The \mpo{overlap between the BI and HE samples of electrons reflects the importance of SSA by Buneman waves for the formation of nonthermal spectral} tails. We find only 6\% of BI electrons in the HE electron sample, \mpo{about three times the fraction of chance coincidences.}
Thus, only a small portion of electrons pre-accelerated in the Buneman instability region finally appears in the nonthermal spectral tail downstream of the shock. 
We must conclude that the Buneman instability does not play a significant role in the formation of the nonthermal tail \jn{for the} case of $m_i/m_e=200$, \mpo{and in any case the tail is weak.}
The shock drift acceleration observed in \cite{Matsumoto2017} is not seen in our simulations. There are two possible explanations: \mpo{the 2D character of our simulations or the} strictly perpendicular geometry of our shock setup.

\subsection{The Influence of SSA on Nonthermal Downstream Population} \label{trans-coef}



\begin{figure*}[!t]
\centering
\includegraphics[width=0.49\linewidth]{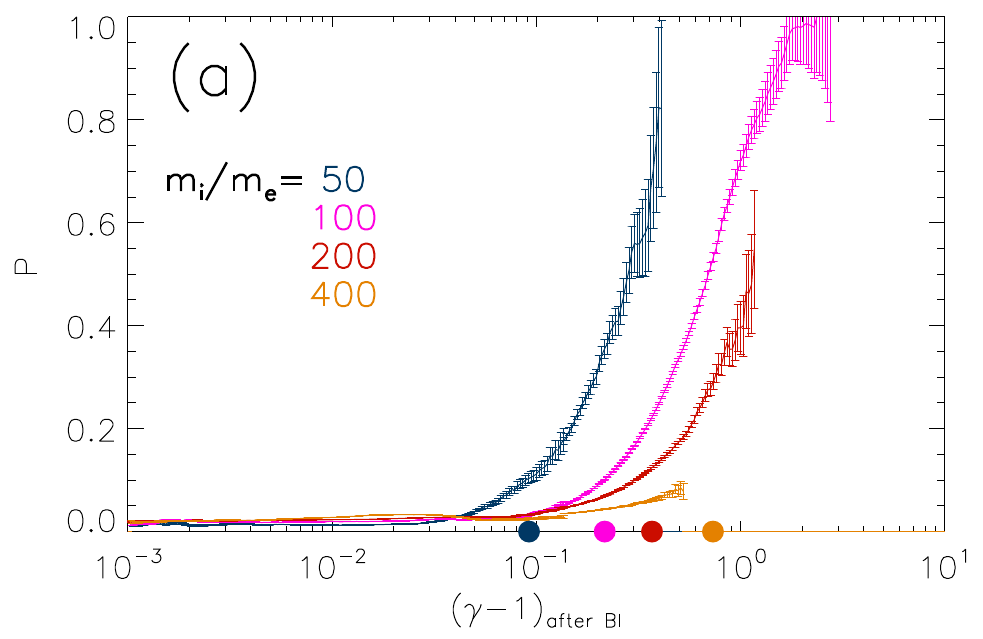}
\includegraphics[width=0.49\linewidth]{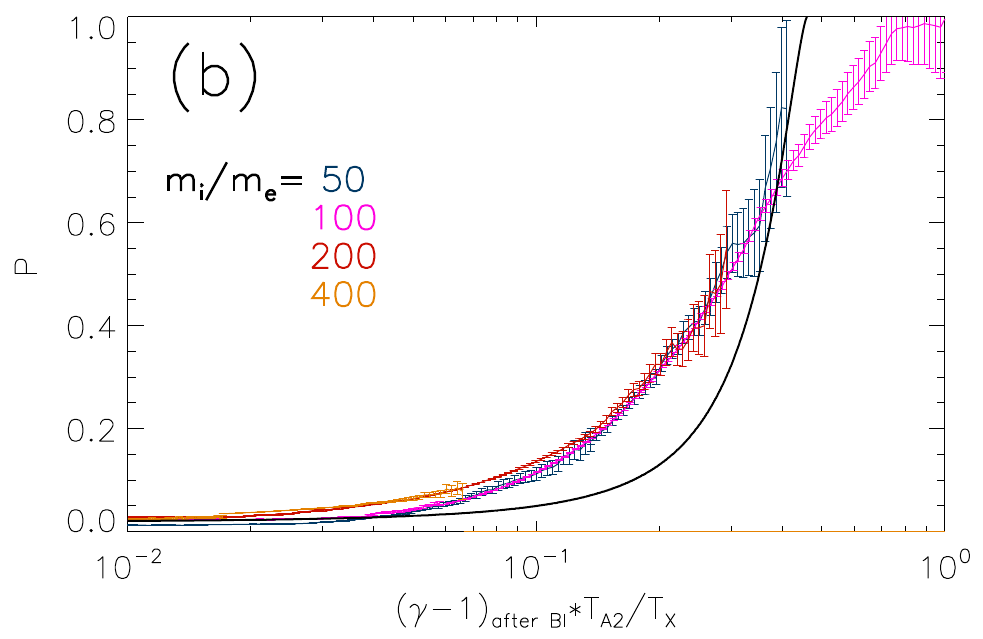}
\caption{Panel (a) -- the probability, P, to reach $E> 5\,k_B T$ after pre-acceleration via SSA in the Buneman instability region. 
The blue line corresponds to run A2 ($m_i/m_e=50$), \rev{magenta} line -- run C2 ($m_i/m_e=100$), red line -- run E2 ($m_i/m_e=200$), and orange line -- run F2 ($m_i/m_e=400$). Circles on the $x$-axis indicate the downstream thermal energy for each run. Panel (b) -- the same plot as in panel (a), 
\rev{but the $x$-axis of each line is rescaled to the downstream temperature of run A2, $T_{A2}$, i.e., by factor $T_{A2}/T_X$, where $T_X$ is the downstream temperature for runs $X=\rm {A2}$, C2, E2 and F2}.
The black solid line designates $P_\mathrm{est}$ calculated via Equation~\ref{fakeP}.}
\label{ele_trans_thres1}
\end{figure*}

\jn{In the last section} \mpo{we established that nonthermal spectral tails are mainly produced during particle interactions with magnetic turbulence in the shock ramp and at the overshoot, and that only a small fraction of electrons pre-accelerated via SSA in the Buneman instability region is still suprathermal in the downstream region.} To further quantify the role of SSA in electron injection,  
let us select electrons from the region right behind the leading edge of the shock foot, i.e., particles that have just crossed the Buneman wave region, and \mpo{ask which part of them eventually reaches a high energy. }
Figure~\ref{ele_trans_thres1}a shows \mpo{as function of the initial energy (after SSA)} the probability, $P$, for an electron \mpo{to have $E > 5k_BT$ in the downstream region}, calculated for runs A2, C2, E2 and F2. 
$P$ increases with the initial energy in all runs, \mpo{and the data points terminate at the maximum energy reached by SSA for the run in question.} \ab{The green line extends up to  $(\gamma-1)\approx 3$ because the strongest SSA is observed in run C2.
As expected, $P$ is about 1.85\% for low-energy electrons ($(\gamma-1)\lesssim 0.05$) in all runs, because all low-energy electrons have \mpo{the same odds} to be accelerated up to nonthermal energies.
Electrons with energy smaller than the downstream thermal energy, $k_BT/m_{e}c^2$,  indicated by colored circles in Fig.~\ref{ele_trans_thres1}a, rarely reach the high-energy tail of the downstream energy distribution.
Electrons with energy $(\gamma-1)_\mathrm{after \  BI}>k_BT/m_{e}c^2$ have \mpo{at least a $10\%$ chance} to reach the high-energy tail. If an electron is accelerated \mpo{by SSA to $(5-10)k_BT/m_{e}c^2$, then it will survive and be in the high-energy tail of the downstream spectrum as well.}}

\mpo{The apparent difference of the probability, $P$, to have a high energy in the four simulations entirely reflects the variation in mass ratio. For a given shock speed the mass ratio determines the downstream temperature of the plasma. Rescaling the initial energy to the downstream temperature of run A2 as reference, i.e., defining the normalized initial energy $(\gamma -1)\,T_{A2}/T_{X}$ for run $X$, we note from Figure~\ref{ele_trans_thres1}b that for all runs the curves of $P$ overlap within errors. The same applies to shocks with cold upstream plasma (runs A1, C1, E1, F1) and runs B and D. It hence appears to be universally so that there is strong electron energization in the shock ramp and at the overshoot that is largely independent of the energy electrons might gain by SSA in the Buneman zone. }

\mpo{Inspection of electron energy spectra demonstrates that most electrons pass through the Buneman waves without significant interaction; only minor heating is observed. A small fraction of electrons, approximately $0.5\%$, experience a strong energy boost.
The average energy gained through SSA, roughly the average energy of electrons with $(\gamma -1)> 0.1$, is about $\langle\gamma -1\rangle\simeq 0.155$ for all runs considered here. It slightly depends on the fulfillment of the trapping condition (Eq.~\ref{trapping} or~\ref{trappingnew}); the energy gain is the smallest in run D and the largest for run C, but all values fall in range of $\langle\gamma -1\rangle = 0.146-0.163$. For runs with the same compliance of the trapping conditions, the energy boost is independent of the mass ratio. } 

\mpo{The downstream electron temperature scales almost linearly with the mass ratio (see Table~\ref{table-spectra2}); in fact approximately $5\%$ of the upstream ion energy is transferred to the downstream electrons, $k_BT \simeq 0.05 \,m_i v_{sh}^2/2$. Therefore only for small mass ratios, e.g., $m_i/m_e=50$, the energy gain by SSA is significant compared to that accumulated at the ramp and the overshoot, and a noticeable fraction of SSA-generated high-energy electrons are still in the high-energy tail of the downstream spectra. A consequence is the anticorrelation between NTEF and the downstream temperature discussed in Section~\ref{downstream}. More specifically, the fraction of SSA-generated high-energy electrons that are still in the high-energy tail of the downstream spectra is roughly $ 30\%$ in run A ($\mi/\me=50$), $10\%$ in runs B and C ($\mi/\me=100$), $ 5\%$ in runs D and E ($\mi/\me=200$), and $0.3\%$ in run F ($\mi/\me=400$). Combining that with the share of electrons that do receive a boost by SSA for runs A, B, E and F, which satisfy the trapping condition (Eq.~\ref{trapping}), about $0.5\%$, we find the NTEF that can be attributed to SSA in the Buneman region as $0.15\%$ for run A, $0.05\%$ for run B, $ 0.025\%$ for run E, and $0.0015\%$ for run F. Only for run A the result is not compatible with zero within the uncertainties (see Table~\ref{table-spectra2}). }

\mpo{The question is whether or not energy gain by SSA and energization in the Weibel instability zone are independent of each other. As very simple test we devise a model in which the two energy gains are simply added. That of the Weibel zone is described by a Maxwellian of temperature $k_BT$, to which we add as offset the energy gain by SSA. We thus calculate the expected probability to find an electron at $E > 5\,k_BT$ as integral over a correspondingly shifted Maxwellian,
\begin{align}
P_\mathrm{est} &\left(y=\frac{(\gamma -1)_\mathrm{after\ BI}\,m_e c^2}{k_B T}\right) \nonumber\\
 &= \frac{2}{\sqrt{\pi}}\ \int_5 dx \sqrt{x+y} \,\exp\left(-\left[x+y\right]\right) ,
\label{fakeP}
\end{align}
which is added as black solid line to Figure~\ref{ele_trans_thres1}b. Clearly, this simple model is a poor fit to the data, and it is evident that the energy gain by SSA has on average a enhancing impact on the subsequent energization in the Weibel zone, otherwise the observed probability, $P$, would not exceed the estimate, $P_\mathrm{est}$, in the range of average SSA gain, $\langle\gamma -1\rangle\simeq 0.15$. The impact is not strong enough to render SSA the dominant process though.}


\mpo{
The energy per electron available for acceleration is the beam energy of upstream electrons relative to the shock-reflected ions~\citep{2009PhPl...16j2901A},
\begin{equation}
E_\mathrm{avail}\approx 2\,\me\,\vsh^2 \simeq 0.15\,m_e c^2\ ,
\label{eq:enerPa}
\end{equation}
where the numerical value applies to simulations with in-plane configuration. 
In principle this energy should turn into electron heat until the thermal velocity spread becomes comparable with the relative velocity and the Buneman waves can no longer grow. After passing the Buneman-wave zone, a spectral tail is indeed observed (Panel b2 of Figure~\ref{ele_acc_HL1_fig}), but the average energy of electrons in the tail is commensurate with Equation~\ref{eq:enerPa} and only few electrons populate it. Hence, only a small fraction of the available energy is transferred to electron heat, about $0.6\%$ in case of in-plane simulations. For the out-of-plane configuration, which better represents the 3D situation, about $7\%$ of the electrons are heated to on average $\langle\gamma -1\rangle \approx 0.31 $ (see Paper I), which for the slightly higher shock speed $\vsh=0.294 c$ yields an energy transfer efficiency of $13\%$. This conversion rate is independent of the mass ratio, but does depend on compliance with the mass-ratio dependent trapping conditions.}

\mpo{We established that, compared to the heating in the Weibel zone, i.e., at the ramp and the overshoot, SSA provides relatively little energization, and that weakness becomes worse with increasing mass ratio, $\mi/\me$. This complicates the extrapolation of simulation results for small mass ratio to real systems. As the post-shock electron temperature, $T_e$, increases with mass ratio, for the realistic mass ratio SSA might play a small role in populating the nonthermal tail in the downstream region, unless much stronger Buneman waves than seen in the simulations here are present to produce energetic electrons with $(\gamma-1) \gtrsim 3\, k_B T_e/m_ec^2$, for which Figure~\ref{ele_trans_thres1}b indicates a reasonable chance, $P\approx 30\%$, for the preheated electrons to be in the spectral tail downstream of the shock.}

As noted in Section~\ref{introduction}, 2D simulations of perpendicular shocks performed with out-of-plane magnetic field geometry well capture the Buneman instability and electron SSA, whereas 2D in-plane configurations \mpo{provide a realistic 3D picture} of the Weibel instability region and the second-order Fermi-like acceleration processes. \ab{We have two simulations \mpo{that differ in the magnetic-field configuration but otherwise have} almost the same physical parameters ($m_i/m_e=100$ and $\ma \approx 32-36$), namely run B with in-plane configuration and run G from Paper I with out-of-plane configuration. \mpo{To estimate the fraction of nonthermal electrons produced by SSA in realistic 3D conditions, one needs to combine spectral information of electrons occupying the Buneman wave region for a run with out-of-plane configuration (the red line in Fig.~6 of Paper I) and the value of $P$ for run B2 which mimics an interaction of SSA-pre-accelerated electrons with the} Weibel instability region and the shock overshoot. The estimated fraction of SSA-pre-accelerated electrons reaching $5k_BT$ in energy is $20\%$. Taking into account the fraction of pre-accelerated electrons in the out-of-plane case, about $7\%$, one finds \mpon{SSA pre-acceleration accounting for an NTEF of roughly $1.5\%$,} and a total NTEF of about $2\%$ for 3D simulations of perpendicular shock with mass ratio $m_i/m_e=100$. \mpo{Hence an additional $0.5\%$ are produced through other mechanisms beyond SSA,}} \ab{which approximately equals the fraction of nonthermal electrons for runs B, \mpon{that to only 10\% of the NTEF arise from} SSA. }

A 3D simulation with oblique magnetic-field configuration~\citep{Matsumoto2017} and comparable mass ratio produces about 1.5\% of nonthermal electrons, which is a little bit \mpon{below} our estimate. 
This suppression of nonthermal fraction can be explained by the 30\% higher temperature of the downstream electrons in 3D simulations compared to the corresponding 2D simulations with in-plane magnetic-field configuration, if we assume the same anticorrelation between downstream electron temperature and nonthermal electron fraction that we observe in our 2D in-plane simulations with the same mass ratio.
However, \mpon{oblique 3D shocks allow for at least one additional acceleration process, namely stochastic shock drift acceleration (SSDA), to operate} in the shock ramp~\citep{Matsumoto2017}. This process accelerates electrons to \mpon{higher energies than is possible with heating or acceleration via chaotic Fermi-like process at perpendicular 2D shocks, and it finally produces a power-law spectrum of electrons downstream of the shock.} The maximal achievable energy is higher by a factor of a few.
We should note that the dependence on the mass ratio of the electron heating in the shock transition may affect the efficiency of SSDA. \mpon{The electron spectra downstream of oblique shocks should be investigated with} 2D or 3D simulations with larger mass ratios.

\section{Summary and discussion} \label{summary}

This work is the second paper in a series \mpo{investigating} electron injection processes at nonrelativistic perpendicular collisionless shocks with high Alfv\'enic Mach numbers by means of 2D3V numerical PIC simulations. Our previous studies indicated \mpo{that SSA operating at the shock foot is} a first-stage electron pre-acceleration mechanism. It was shown that SSA works provided that the Alfv\'enic Mach number \mpo{exceeds a certain minimum} (the trapping condition), which was demonstrated with 2D simulations \mpo{with} out-of-plane configuration of the mean upstream magnetic field component \citep{2012ApJ...755..109M}. In \cite{2015Sci...347..974M} and \cite{2017ApJ...847...71B} we showed that in 2D simulations that use a field component which lies in the simulation plane, SSA is followed by additional non-adiabatic acceleration in the shock ramp. In Paper I we demonstrated that with an in-plane magnetic-field configuration we cannot achieve the same SSA efficiency as in simulations with out-of-plane magnetic field or 3D simulations.
\mpo{We note} that much of this difference results from significantly \mpo{faster streaming} between reflected ions and upcoming electrons in the out-of-plane case, \mpo{accounting for which requires a} significant modification of the simulation setup. 
Here we further investigate electron acceleration processes at perpendicular high-$\ma$ shocks using 2D setups with in-plane magnetic field configuration.
In particular, we are interested in the \mpo{level of the initial electron energization via SSA and its impact on the final downstream spectra, and how this role of SSA scales with the mass ratio and the Alfv\'enic Mach number.}

Our results can be summarized as follows:


\begin{itemize}

\item 
There is no strict correlation between the SSA efficiency in the shock foot and the nonthermal-electron fraction \mpo{downstream of the shock. This suggests that SSA is not the only relevant process for the generation of suprathermal electrons and their injection into DSA. }

\item \mpo{For the mass ratio $m_i/m_e=200$, only $6\%$ of the electrons pre-accelerated via SSA are still in the high-energy tail in the downstream spectra. The likelihood of an electron being nonthermal in the downstream region depends in a unique way on the energy it had after passing the SSA zone, normalized to the downstream electron temperature. This indicates that for the realistic mass ratio SSA has a minor influence on the formation of nonthermal electron in the downstream region. SSA-energized electrons gain energy in the ramp/overshoot regions with a similar rate as does the bulk of electrons. The electrons that end up in the nonthermal tail downstream gained energy at unusually high rate, suggesting that they were at the right place at the right time to be efficiently energized in the shock ramp by a second-order Fermi process or via interaction with magnetic-reconnection sites.}

\item \mpo{The average energy gain by SSA is commensurate with the mean energy per electron available for driving Buneman waves, and it does not depend on the mass ratio used in the simulation. It does depend on the magnetic-field configuration though and is twice as large for out-of-plane setups than for in-plane simulations. The fraction of electrons that are trapped by the Buneman waves and undergo SSA is much smaller than $100\%$, and so there is no energy concentration on the relatively few electrons that experience SSA. The downstream electron temperature depends on the mass ratio, and so for high mass ratios the energy gain in the ramp and at the overshoot becomes larger than that attainable with SSA.
Our simulations hence demonstrate that the dominant acceleration takes place in the turbulent shock ramp, and only a small fraction of electrons pre-accelerated via SSA in the shock foot experiences further energization in the ramp.} 

\end{itemize}



\ta{Our conclusions are valid for nonrelativistic shocks with shock speed  $v_{\rm sh}=0.263c$. A realistic shock speed is ten times smaller, namely $v_{\rm sh}\approx 8000$~km/s, and the situation could be different. We have shown that the downstream electron temperature is roughly proportional to the ion flow kinetic energy, $T_e \propto 1/2 m_i \vsh^2$, which is at most marginally relativistic $T_e \lesssim m_e c^2$ for a realistic shock speed. As long as the trapping condition is satisfied, SSA may potentially provide non-thermal electrons of comparable, mildly relativistic energies even for a realistic shock speed. If this is the case, SSA may remain important for injection. However, the scaling of SSA efficiency with the shock speed is not entirely clear at this point.}

\mpo{We should note that pre-acceleration to a few times the downstream temperature is likely insufficient for injection into DSA. At least in its standard form DSA builds on the assumption that particles can freely pass through the shock. As the shock thickness is commensurate with the ion Larmor radius, and avoiding scattering in the shock transition mandates that the electron Larmor radius be much larger than the shock thickness, we arrive at a requirement \art{for the electron momentum, $p_\mathrm{e} \gg \mi \vsh\approx 25\ \mathrm{MeV}/c$ for $\vsh=8000\ \mathrm{km/s}$. Hence being suprathermal is not enough.}}

\mpon{3D simulations of oblique shocks demonstrate the operation of an additional efficient acceleration mechanism, SSDA, which may accelerate electrons to considerably higher energies \citep{Matsumoto2017}.} \ab{At oblique shocks SSA plays a critical role in reflecting electrons back into the foreshock \citep{2007ApJ...661..190A} and influences the injection of electrons into SSDA, and so our conclusions apply only to strictly perpendicular shocks. }

\ab{This paper is the second of a series investigating different aspects of electron acceleration processes at non-relativistic perpendicular shocks using PIC simulations. On account of the costliness of 3D simulations, a number of 2D3V simulations of quasi-perpendicular shocks with different mass ratios, Mach numbers and shock velocities are conducted to investigate \ab{the scaling behavior of acceleration processes such as SSA, stochastic Fermi-like processes, magnetic reconnection, shock drift acceleration, etc.} Magnetic reconnection in the shock transition, plasma heating, and the generation of turbulent magnetic field will be covered in forthcoming publications.}

\acknowledgments


The work of J.N. has been supported by Narodowe Centrum Nauki through research project DEC-2013/10/E/ST9/00662. This work was supported by JSPS-PAN Bilateral Joint Research Project Grant Number 180500000671.
The numerical experiment was possible through a 10 Mcore-hour allocation on the 2.399 PFlop Prometheus system at ACC Cyfronet AGH. Part of the numerical work was conducted on resources provided by the North-German Supercomputing Alliance (HLRN) under projects bbp00003 and bbp00014.

\bibliographystyle{apj}
\bibliography{ref}

\end{document}